# A Numerical Treatment of the Rf SQUID: I. General Properties and Noise Energy

Reinhold Kleiner[1], Dieter Koelle[1] and John Clarke[2]

[1]Physikalisches Institut-Experimentalphysik II, Universität Tübingen, 72076 Tübingen, Germany

[2]Department of Physics, University of California, Berkeley, California 94720-7300,

and

Materials Sciences Division, Lawrence Berkeley National Laboratory, Berkeley, California 94720

E-mail: kleiner@uni-tuebingen.de

*We investigate the characteristics and noise performance of rf Superconducting Quantum Interference Devices (SQUIDs) by solving the corresponding Langevin equations numerically and optimizing the model parameters with respect to noise energy. After introducing the basic concepts of the numerical simulations, we give a detailed discussion of the performance of the SQUID as a function of all relevant parameters. The best performance is obtained in the crossover region between the dispersive and dissipative regimes, characterized by an inductance parameter $\beta'_L \equiv 2\pi L I_0/\Phi_0 \approx 1$; L is the loop inductance, $I_0$ the critical current of the Josephson junction, and $\Phi_0$ the flux quantum. In this regime, which is not well explored by previous analytical approaches, the lowest (intrinsic) values of noise energy are a factor of about 2 above previous estimates based on analytical approaches. However, several other analytical predictions, such as the inverse proportionality of the noise energy on the tank*

*circuit quality factor and the square of the coupling coefficient between the tank circuit and the SQUID loop, could not be well reproduced. The optimized intrinsic noise energy of the rf SQUID is superior to that of the dc SQUID at all temperatures. Although for technologically achievable parameters this advantage shrinks, particularly at low thermal fluctuation levels, we give an example for realistic parameters that leads to a noise energy comparable to that of the dc SQUID even in this regime.*

# 1. INTRODUCTION

For more than 30 years, rf Superconducting Quantum Interference Devices (SQUIDs) have served as reliable and sensitive detectors of magnetic flux. Given the voluminous literature on the subject, one might have expected that all important features of this device had been treated in detail. Although this is certainly true for many aspects, it appears, however, that detailed *numerical* simulations solving the underlying dynamic Langevin equations have not yet been performed. In contrast, such simulations have been standard for the analysis of dc SQUIDs since the mid 1970's [1,2,3]. Early theoretical (and also experimental) investigations of the noise performance of rf SQUIDs focused on the so-called dissipative regime in which the total magnetic flux $\Phi$ enclosed in the SQUID loop is a multivalued function of the applied flux. In this regime, which is characterized by an inductance parameter $\beta'_L \equiv 2\pi L I_0 / \Phi_0 > 1$ (where $L$ is the inductance of the SQUID ring, $I_0$ is the junction critical current, and $\Phi_0 \equiv h/2e$ is the flux quantum), the rf drive causes the SQUID to switch between quantum states and to dissipate energy at a rate that is periodic in $\Phi_{ext}$. Between 1972 and 1975, Kurkijärvi [4,5], Kurkijärvi and Webb [6] and Jackel and Buhrman [7] developed a theory for the flux noise of the rf SQUID that was based on the use of suitable probability functions describing the quantum transition of the SQUID. The analyses included the effect of fluctuations in the SQUID loop, the tank circuit and the preamplifier; the last often limits the performance of rf SQUIDs cooled to liquid helium temperatures. A detailed analysis and optimization of all forward and reverse interactions between the rf SQUID input and its output were given by Ehnholm[8] in 1977. Thorough comparisons between theory and experiment were performed by Giffard and Hollenhorst [9,10].

In the mid 1970's the dispersive regime (characterized by $\beta'_L < 1$), where $\Phi$ vs. $\Phi_{ext}$ is a single-valued function, was addressed by Hansma [11,12], who calculated the SQUID characteristics at zero temperature in the limit $\beta'_L \ll 1$. This approximation was later extended to $\beta'_L \leq 1$ by Rifkin *et al.* [13] and by Sorensen [14]. In the limit $\beta'_L \ll 1$ the effects of small thermal fluctuations were addressed by Likharev and Ulrich [15] and by Danilov and Likharev [16]. The noise performance of both low- and high-inductance interferometers was studied intensively by Danilov, Likharev and Snigirev [17]. While the work mentioned above treated thermal noise as a small perturbation, in 1998 Chesca addressed the case of large fluctuations in the extreme dispersive regime $\beta'_L \ll 1$ by analytically solving the non-stationary Smoluchovski equation describing the noise performance of the rf SQUID [18]. In the limit of very large fluctuations Chesca also gave analytical solutions for the case $\beta'_L \leq 3$. The optimal noise energy can be as low as [17] $\varepsilon \approx 0.47 \cdot [1 + T_T / T\alpha^2 Q\beta'_L f_d] \cdot k_B T / \beta'_L f_c$ at low temperatures ($\Gamma\beta'_L \ll 1$)), increasing as $\exp(\Gamma\beta'_L)$ when fluctuations become large [18]. Here, $f_c = I_0 R / \Phi_0$ is the junction characteristic frequency, $R$ is the junction resistance and $\Gamma = 2\pi k_B T / I_0 \Phi_0$ is the noise parameter. The temperatures $T$ and $T_T$ refer to the SQUID and tank circuit, respectively, and $\alpha = M / \sqrt{LL_T}$ is the coupling parameter between the SQUID and the tank circuit inductance $L_T$; $M$ is the mutual inductance. Finally, $Q$ is the quality factor of the tank circuit, and $f_d$ is the drive frequency. Chesca [18] further pointed out that the factor $(\alpha^2 Q\beta'_L f_d)^{-1}$ is usually much larger than unity and thus dominates $\varepsilon$. In addition, the noise energy in the dissipative regime is worse than in the dispersive regime by a factor of $f_c / f_d \gg 1$; thus, the dispersive regime seems by far superior [18].

These analytical approaches for dispersive rf SQUIDs assume that the tank circuit oscillations are quasi-sinusoidal; this is likely to be valid for weak coupling ($\alpha \ll 1$) and small values of $\beta'_L (\ll 1)$. On the other hand, the transition regime between the dispersive and the dissipative mode, in particular, is not well explored. Here, it appears that numerical simulations are required, and are the subject of this paper. We shall see that indeed very low values of ε can be obtained near $\beta'_L = 1$, remaining low for values of $\Gamma$ as high as unity. To obtain very low values of $\varepsilon$ the coupling coefficient α is required to be well above 0.2.

The remainder of this paper is organized as follows. Chapter 2 addresses the model and explains numerical details. In Chapter 3, we choose sample parameters and discuss current-voltage characteristics and the transfer function, and, in Chapter 4, we use these results to calculate the voltage and noise energy. In Chapter 5 we discuss the optimized circuit by varying a large number of parameters to minimize the noise energy. Chapter 6 contains our conclusions. Appendix A contains a list of symbols used in the paper. We address the noise temperature of rf SQUIDs used as amplifiers in the following paper[19].

## 2. MODEL AND NUMERICAL DETAILS

To model the rf SQUID and its readout we consider the circuit shown in Fig. 1. The rf SQUID consists of a superconducting loop with inductance $L$ coupled to a tank circuit via a mutual inductance $M$. The Josephson junction incorporated into the SQUID loop is described via the resistively and capacitively shunted junction (RCSJ) model [20,21] by a Josephson current with maximum amplitude $I_0$ in parallel with a resistor $R$ and a capacitor $C$. The tank circuit consists of an inductor $L_T$, a capacitor $C_T$ and a resistor $R_T$ and is biased with an oscillating current $I_T = I_d \cos\omega_d t$. The current flowing through $L_T$ is denoted by $I_1$, and that through the right arm by $I_2$. Counterclockwise currents give positive flux. The thermal noise currents arising from the resistors $R$ and $R_T$ are modelled by two independent current sources $I_N$ and $I_{N,T}$ connected in parallel with $R$ and $R_T$, respectively, each with a white spectral density. We allow $R$ and $R_T$ to have different temperatures so that the spectral densities of the current noise produced by $R$ and $R_T$ are $4k_BT/R$ and $4k_BT_T/R_T$, respectively.

We first derive the differential equations describing the circuit of Fig. 1 in dimensioned units. The Josephson junction is described by

$$I_0 \sin \delta + \frac{\Phi_0}{2\pi R}\dot{\delta} + \frac{\Phi_0}{2\pi} C\ddot{\delta} + I_N = J. \quad (1)$$

Here, $\delta$ is the gauge invariant phase difference between the superconducting order parameters on either side of the barrier and $J$ is the current circulating in the SQUID loop. Dots denote time derivatives.

The total flux $\Phi$ through the SQUID loop is connected to the gauge invariant phase difference $\delta$ via $\delta = -2\pi\Phi/\Phi_0$, yielding $\Phi = \Phi_{ext} + LJ + MI_1$. Here, $\Phi_{ext}$ is the external (static) flux. For the tank circuit the voltages $U_{LT}$, $U_{RT}$ and $U_{CT}$ across $L_T$, $R_T$, and $C_T$, respectively, are related by $U_{LT} = U_{CT} + U_{RT}$. One way to proceed would be to replace the time derivatives of these voltages by the currents through the corresponding elements, for example, yielding a second order differential equation in $I_1$. In this formulation, however, the first time derivative of the current through $R_T$ appears, making the numerical treatment of noise inconvenient. It is thus better to use the flux $\Phi_T$ through $L_T$ as a variable. With $L_T I_1 + MJ = \Phi_T$, $U_{LT} = \dot{\Phi}_T$, $I_T = I_1 + I_2$, $I_2 = C_T \dot{U}_{CT} = (U_{RT} + U_{NT})/R_T = U_{RT}/R_T + I_{NT}$, where $U_{NT}$ and $I_{NT}$ denote the noise voltage across and noise current through the resistor $R_T$, we find

$$I_T = I_1 + C_T \dot{U}_{CT}. \quad (2)$$

Here, $I_T = I_d \cos\omega_d t$, $I_1 = [\Phi_T - MJ]/L_T = [\Phi_T + \alpha\gamma_L(\Phi_0\delta/2\pi + \Phi_{ext})]/(1-\alpha^2)L_T$ and $J = [-\Phi_0\delta/2\pi - \Phi_{ext} - \alpha\Phi_T/\gamma_L]/(1-\alpha^2)L$. We defined the parameter $\gamma_L = \sqrt{L_T/L}$ which, as we see below, acts as a scaling parameter for voltages, currents and fluxes in the tank circuit. Further, from the current $I_2$ through the right arm of the tank circuit, we have

$$\dot{U}_{CT} = \frac{1}{R_T C_T}(U_{RT} + U_{NT}) = \frac{1}{R_T C_T}(\dot{\Phi}_T - U_{CT} + U_{NT}) \ . \tag{3}$$

Equations (1) to (3) form a system of coupled differential equations for the variables $\delta$, $\Phi_T$ and $U_{CT}$ that we can use to calculate the dynamics of the rf SQUID. Before doing so, however, we convert the relevant equations into a dimensionless notation. We normalize currents $i$ in units of $I_0$, fluxes $\varphi$ in the SQUID loop in units of $\Phi_0$, fluxes $\varphi_T$ in the tank circuit in units of $\Phi_0/2\pi$, voltages $u$ in units of $I_0 R$ and time $\tau$ in units of $\Phi_0/2\pi I_0 R$. We further define the Stewart-McCumber parameter $\beta_c = 2\pi I_0 R^2 C / \Phi_0$ and the reduced inductance $\beta'_L = 2\pi I_0 L / \Phi_0$. With these parameters for the rf SQUID loop, we find

$$\beta_c \ddot{\delta} + \dot{\delta} + \sin\delta + i_N = j = \frac{-1}{1-\alpha^2}\left[\frac{1}{\beta'_L}(\delta + 2\pi\varphi_{ext}) + \alpha\frac{\varphi_T}{\beta'_L \gamma_L}\right] \ . \tag{4}$$

For the tank circuit, with $\varphi_T = 2\pi\Phi_T/\Phi_0$, we obtain

$$i_d \cos f_d \tau = \frac{\varphi_T}{\beta'_L \gamma_L^2} - \alpha\frac{j}{\gamma_L} + \frac{1}{f_0^2 \beta'_L \gamma_L^2}\dot{u}_{cT} \tag{5}$$

and

$$\dot{u}_{cT} = f_0 Q_0(\dot{\varphi}_T - u_{cT} + u_{NT}) \ . \tag{6}$$

Here, $Q_0 = \sqrt{L_T/C_T}/R_T$ denotes the (unloaded) quality factor and $f_0 = \Phi_0/2\pi I_0 R\sqrt{L_T C_T}$ the (unloaded) resonance frequency in units of the junction characteristic frequency $f_c = I_0 R/\Phi_0$. It is convenient to define the resistance ratio $\gamma_R = R_T/R$ which can be expressed as $\gamma_R = \beta'_L f_0 \gamma_L^2 / Q_0$.

We next give explicit expressions for the noise currents and voltages. The functions $i_{N,0}$ and $i_{NT,0}$ consist of a sequence of Gaussian distributed random numbers with zero average and a mean square deviation of unity. The random numbers change after a time interval $\tau_{noise}$ well below any response time of the system (typically we take $\tau_{noise} = 0.5$). We express the noise current through the junction resistor as $i_N = \sqrt{2\Gamma/\tau_{noise}} \cdot i_{N,0}$. Similarly, for the noise voltage $u_{NT}$ across $R_T$ we obtain

$$u_{NT} = \gamma_R \cdot i_{NT} = \gamma_R \cdot \sqrt{2\Gamma_T/(\tau_{noise}\gamma_R)} \cdot i_{NT,0} = \sqrt{\beta'_L f_0 \gamma_L^2 / Q_0} \cdot \sqrt{2\Gamma_T/\tau_{noise}} \cdot i_{NT,0},$$

with $\Gamma_T = 2\pi k_B T_T / I_0 \Phi_0$.

The above differential equations (4) – (6) depend on the three variables $\delta$, $u_c$ and $\varphi_T$ and on the 11 parameters $\gamma_L, \beta'_L, \beta_c, \alpha, Q_0, f_0, \Gamma, \Gamma_T, f_d, i_d$ and $\varphi_{ext}$. An important quantity obtained directly from these equations is the voltage across the tank circuit, which can be expressed as $u_T = u_{LT} = \dot{\varphi}_T$. From $u_T$ we obtain a normalized "static voltage" $v_T$ across the tank circuit as follows. First, after we have changed some parameters of the calculation, we allow the system to relax for about 1000 periods of ac drive. After this relaxation step a time trace of $u_T$ is taken over typically 100 - 1000 periods of the rf drive. This time trace is subsequently Fourier transformed and the Fourier channel corresponding to the drive frequency $f_d$ is used as the "dc voltage" $v_T$. Note that for a time trace of 100 - 1000 periods of the rf drive one channel of the Fourier transform corresponds to a frequency interval $\Delta f$ between $10^{-2}$ and $10^{-3} f_d$. Using this channel as the output assumes that we have a detector to measure the amplitude of $u_T$ at the drive frequency in the (narrow) bandwidth $\Delta f$. The voltage $v_T$ serves as the low-frequency voltage output, to be plotted as a function of bias current, applied flux etc. In order to calculate low frequency fluctuations a series of 2$N$ values of $v_T$ (typically $N$ = 1024) is used to constitute a time trace $v_T(\tau)$, to be Fourier transformed to obtain the voltage noise power $s_{vT}$ (in units of $I_0 R \Phi_0 / 2\pi$), the flux noise $s_\varphi = s_{vT}/(dv_T/d\varphi_{ext})^2$ and the normalized noise energy $e = \pi s_\varphi / 2\Gamma \beta'_L$. This definition ensures that the units of $e$, $2\Phi_0 k_B T / I_0 R$, are the same as that used for dc SQUIDs [22], allowing a direct comparison of the two devices.

A major task is to find the parameters that minimize $e$. One faces a high dimensional optimization problem that perhaps has been the main obstacle to addressing the rf SQUID numerically long ago. Fortunately, at least one of the parameters, $\gamma_L$, appears as a simple scaling parameter. It turns out that voltages, fluxes and currents across the tank circuit can be obtained from the case $\gamma_L = 1$ by dividing voltages and fluxes by $\gamma_L$ and multiplying the currents with $\gamma_L$. One can see this by introducing $\tilde{u}_{cT} = u_{cT}/\gamma_L$, $\tilde{\varphi}_T = \varphi_T/\gamma_L$, $\tilde{i} = i\gamma_L$ and replacing (4) – (6) with

$$\beta_c \ddot{\delta} + \dot{\delta} + \sin\delta + i_N = j = \frac{-1}{1-\alpha^2}\left[\frac{1}{\beta'_L}(\delta + 2\pi\varphi_{ext} + \alpha\tilde{\varphi}_T)\right] \quad (7)$$

$$\tilde{i} = \frac{\tilde{\varphi}_T}{\beta'_L} - \alpha j + \frac{1}{f_0^2 \beta'_L}\dot{\tilde{u}}_{cT}, \quad (8)$$

and

$$\dot{\tilde{u}}_{cT} = f_0 Q_0 (\dot{\tilde{\varphi}}_T - \tilde{u}_{cT} + \tilde{u}_{NT}). \quad (9)$$

Below we thus consider only the case $\gamma_L = 1$ unless otherwise stated. We note that the static voltage $v_T$ across the tank circuit, the transfer function $dv_T/d\varphi_{ext}$ and the amplitude of the voltage noise $s_v^{1/2}$ also scale with $\gamma_L$. By contrast, flux noise $s_\varphi = s_v / v_\varphi^2$ and noise energy $e$ are independent of $\gamma_L$.

To optimize $e$ for a set of parameters (for example $i_d$, $f$, $\varphi_{ext}$, $\alpha$, $\beta'_L$) the first step is to guess an initial set of reasonable parameters and calculate $v_T$, $dv_T/d\varphi_{ext}$ and $s_v$, and from those $s_\varphi$ and $e$. Subsequently the parameters are varied one by one with a suitable step width. After each step the parameter value yielding the lowest value of $e$ is adopted and the next parameter is varied until $e$ has saturated or some maximum iteration number is exceeded. During the variation, depending on whether or not a new value of the parameter reduces $e$, the step width to change this parameter during the next iteration is either increased or decreased by typically a factor of 2. Figure 2 gives an example how $e$ and the varied parameters evolve during the optimization procedure. Fixed parameters for this simulation were $Q_0 = 100$, $\varphi_{ext} = 0.25$, $f_0 = 0.1$, $\beta_c = 0$ and $\Gamma = \Gamma_T = 0.025$. After 5 steps $e$ already reached a level of about 0.6. With subsequent iterations it decreased only slowly. Furthermore, the parameters varied reached almost constant values after about 15 iterations. As we see later, for the above fixed parameters, the value $e \approx 0.5$ is reached for a coupling parameter $\alpha$ close to 1. The optimization procedure of Fig. 2, which was terminated after 15 iterations, thus would have yielded a noise energy that is about 20 % too large, with $\alpha \approx 0.5$ well below 1. Consequently, we will have a typical systematic error in $e$ of some 10-20%. We also note that there are many local minima for $e$ that act as potential error sources; for example, as a function of $i_d$, the transfer function and thus also $e$ oscillates. The values given below for optimized noise energies should always be regarded as upper limits. Another issue to address is the extent to which the transfer function rather than $e$ should be optimized to find an optimal bias point. At least for dc SQUIDs, this procedure is frequently used. However, in the calculations for the rf SQUID we found that, on the one hand, for some parameters $dv_T/d\varphi_{ext}$ can exhibit jumps or hystereses [for example, see Figs. 4(c) and 5(c)] mimicking infinite transfer functions. On the other hand, we sometimes found that large transfer functions coincide with a large value of $s_v$, leading to a non-optimal value of $e$, while for some other, non-optimal, values of $dv_T/d\varphi_{ext}$ both $s_v$ and $e$ were much lower. We conclude that to find the lowest noise energy this quantity itself needs to be optimized.

## 3. CURRENT-VOLTAGE CHARACTERISTICS AND TRANSFER FUNCTIONS

We begin by comparing the numerically calculated current-voltage characteristics ($i_d$ vs $v_T$) with analytical expressions obtained by Hansma [11] and Chesca [18]. At zero temperature, converted to our units, Hansma's expression is

$$i_d = \frac{\sqrt{1+\xi^2}}{Q_0} \left\{ \left[ \frac{\tilde{A}}{\sqrt{1+\xi^2}} \cos(2\pi\varphi_{ext}) \cdot J_1(\beta'_L \alpha \gamma_L i_T) + \tilde{i}_T \cos\theta \right]^2 + \left( \tilde{i}_T \sin\theta \right)^2 \right\}^{1/2} . \qquad (10)$$

Here, $\tilde{A} = 2Q_0\alpha/\gamma_L$, $\tilde{i}_T = v_T / \beta'_L \gamma_L^2 f$, $\theta = \arctan\xi + \pi/2$, the detuning parameter $\xi = 2Q_0(f_d/f_0 - 1)$ and $J_1(x)$ is the first order Bessel function. Chesca [18] gives several expressions depending on the values of $\beta'_L$ and $\Gamma$. For $\beta'_L \ll 1$, in our units, Chesca obtains

$$i_d = \frac{1}{\beta'_L \gamma_L} \left\{ \left[ \tilde{a}(\frac{1}{Q_0} + \frac{\alpha^2 \omega_R}{1+\omega_R^2}) - \frac{2\alpha\omega_R}{\sqrt{1+\omega_R^2}} F \right]^2 + \left[ 2\tilde{a}(\tilde{\xi} - \frac{\alpha^2 \omega_R}{1+\omega_R^2}) - \frac{2\alpha}{\sqrt{1+\omega_R^2}} F \right]^2 \right\}^{1/2} , \qquad (11)$$

where $\tilde{a} = v_T / f\gamma_L$, $\omega_R = f_d \cdot \beta'_L$, $F = \beta'_L \exp(-\Gamma\beta'_L/2) J_1(a^*)\cos 2\pi\varphi_{ext}$, $a^* = \tilde{a}\alpha/(1+\omega_R^2)^{1/2}$ and $\tilde{\xi} = f_d/f_0 - 1$. Chesca[18] gives another expression for larger values of $\beta'_L$ that is valid for large fluctuations ($\Gamma \gg 1$) only, and that is not reproduced here. In contrast to (10), (11) contains the temperature via the function $F$. In addition, the argument of the Bessel function contains a factor $(1+\omega_R^2)^{-1/2}$ that is absent in (10); for $f_d\beta'_L \ll 1$ this is a small correction. Note that in both (10) and (11) $i_d$ is calculated as a function of $v_T$, while numerically, we calculate the inverse function. Thus, to obtain the transfer function $dv_T/d\varphi_{ext}$ at fixed $i_d$ from (10) or (11) we compute the zeroes of $f(i_d) = i_d(v_T) - i_d$ for given values of $i_d$.

Figure 3(a) shows a comparison of the zero temperature $i_d$–$v_T$ characteristics calculated numerically (solid circles) from (10) (referred to as 'PH' – solid lines) and from (11), (referred to as 'BC' – dashed lines). The curves are for $\varphi_{ext} = 0$ and 0.5. Other parameters are $Q_0 = 100$, $f_0 = 0.1$, $f_d = 0.101$ ($\xi = 2$ or $\tilde{\xi} = 0.01$) and $\alpha = 0.1$. We see that discrepancies between the curves are small for this set of parameters. This also holds for Fig. 3(b) where we plot the transfer function $dv_T/d\varphi_{ext}$ at $\varphi_{ext}$ vs. $i_d = 0.25$ as well as for Fig. 3(c) where we plot $v_T$ vs. $\varphi_{ext}$ for $i_d = 0.8$, that is for a bias near the optimal transfer function. Other values of, for example, the detuning parameter give similarly good agreement. The differences become more severe, however, at larger values of $\alpha$. Figure 4 shows plots of $i_d$ vs. $v_T$, transfer function vs. $i_d$ and $v_T$ vs. $\varphi_{ext}$ for $i_d = 0.25$, organized as in Fig. 3; the parameters are the same except for $\alpha$, which is now 0.3. Although the overall shapes of the numerical and the analytical curves are similar, and, for example, the amplitude of the maximum transfer function is comparable, at a fixed value of $i_d$ both $v_T$ and $dv_T/d\varphi_{ext}$ differ strongly, showing that comparisons between the numerical and analytical curves cannot be made by calculating the SQUID performance simply at a fixed set of parameters. Note that in Fig. 4(c) the BC curve exhibits jumps near $\varphi_{ext} = 0.05$ and 0.95. Also, both analytical $i_d$–$v_T$ characteristics in Fig. 4(a) exhibit regions of negative slope which would result in hystereses in a current biased situation. For other sets of parameters such jumps and hystereses also appear in the numerical calculations, creating regions of parameters to be avoided. One should further note that the values of $Q_0$, $f_0$, $f_d$ and $\alpha$ are not in the regimes where realistic devices are operated. Experimentally, $Q_0$ is typically very large ($\gtrsim 10^5$) while $\alpha$ is extremely small. A value of $f_0 \approx f_d = 0.1$ with a characteristic junction frequency of 100 GHz corresponds to $f_d$ 10 GHz, which is higher than typical rf frequencies used. More realistic values of $Q_0$, $f_0$ and $f_d$ are, however, beyond reach since computation times would become intolerably long.

As a third example Fig. 5 (a) shows $i_d - v_T$ characteristics for $\varphi_{ext} = 0$ and $\varphi_{ext} = 0.5$ for $f_0 = 0.1$, $Q_0 = 100$ and $\Gamma_T = 0.025$ (solid circles). Using the optimization routine, we find that the parameters $f = 0.1066$, $\alpha = 0.725$ and $\beta'_L = 1.21$ give a noise energy $e \approx 0.5$ at $i_d = 0.369$ (see also sections 4 and 5). In this particular optimization $\beta_c = 0$ and $\varphi_{ext} = 0.25$ were kept fixed (we see below that for $\beta'_L \lesssim 1.5$ $e$ does not depend significantly on $\beta_c$ provided the drive frequency is well below the LC resonance of the SQUID loop, which can be expressed as $(\beta'_L \beta_c)^{-1/2}$; also, for $\varphi_{ext}$ near 0.25 $e$ is almost minimal). The $i_d - v_T$ curves at both $\varphi_{ext} = 0$ and 0.25 exhibit several jumps and differ only weakly from the numerical zero temperature calculation for the same parameters, shown by open circles in Fig. 5(a). For comparison, we also display the zero temperature $i_d - v_T$ curve calculated from Hansma's Eq. (10) which, for these parameters, does not differ too much from the numerical curve. As is clearly seen, in a current-biased situation this analytical characteristic is also hysteretic. Figure 5(b) displays the numerical transfer function $dv_T/d\varphi_{ext}$ vs. $i_d$ at $\varphi_{ext} = 0.25$ for $\Gamma = \Gamma_T = 0.025$ (solid circles) and for $\Gamma = \Gamma_T = 0$ (line). The difference between the two curves is barely visible. Prominent features, however, are the large spikes at $i_d = 0.19$ and 0.66. At least at the lower of these values the transfer function may even diverge (the transfer function has been calculated from the difference in $v_T$ at $\varphi_{ext} = 0.25 \pm 0.01$, of course leading to a finite value of $dv_T/d\varphi_{ext}$).

Perhaps suprizingly, the optimal value of $e$ is obtained far from these spikes, namely at the bias indicated by the arrow in Fig. 5(b). Finally, Fig. 5(c) displays $v_T$ vs. $\varphi_{ext}$ for the optimal bias current $i_d = 0.369$. There is a bistable region near $\varphi_{ext} = 0$ and $\varphi_{ext} = 1$. Voltage jumps are indicated by vertical arrows; tilted arrows indicate the direction of flux sweeps. Outside this region, and particularly near $\varphi_{ext} = 0.25$, $v_T$ vs. $\varphi_{ext}$ is continuous, however.

Finally, we demonstrate that the transfer function can be quite robust against noise. Figure 6(a) shows $dv_T/d\varphi_{ext}$ vs. $i_d$ for various values of $\Gamma = \Gamma_T$ (0, 0.025, 0.3, 1.0 and 3.0) and the solid circles in Fig. 6(b) show the transfer function at $i_d = 0.369$ (that is, at the optimal bias for $\Gamma = \Gamma_T = 0.025$) vs. $\Gamma = \Gamma_T$. All other parameters are as in Fig. 5 (b). In (a) the spikes in $dv_T/d\varphi_{ext}$ become strongly suppressed with increasing noise; in addition, note that the first spike shifts to lower values of $i_d$ with increasing noise parameter. The modulus of the transfer function near $i_d = 0.369$ even increases slightly with increasing $\Gamma$ for $\Gamma \leq 1$. It then drops with further increase of $\Gamma$ but still has half of its original value at $\Gamma = 3$. For comparison, the maximum transfer function of a dc SQUID is suppressed by more than an order of magnitude when $\Gamma$ is increased from 0.025 to 1. One thus sees already at this juncture that the noise performance of the rf SQUID degrades much less rapidly than that of the dc SQUID when $\Gamma$ becomes large. The open and grey symbols in Fig. 6(b) show the transfer function vs. $\Gamma$ for $\Gamma_T = 0$ and $\Gamma_T = 10\Gamma$. The curve for $\Gamma_T = 0$ does not differ much from the curve for $\Gamma_T = \Gamma$, showing that the degradation in $dv_T/d\varphi_{ext}$ is dominated by fluctuations in the SQUID loop. The effect of the tank circuit fluctuations becomes more severe for $\Gamma_T = 10\Gamma$, but even then reasonable transfer functions can be obtained at least up to $\Gamma = 1$.

## 4. VOLTAGE NOISE AND NOISE ENERGY: EXAMPLES

We begin this section by looking at sample power spectra $s_{vT}$ and $s_{uT}$ of the (low-frequency) tank voltage $v_T$ and (high-frequency) voltage $u_T$ (Fig. 7). In Fig. 7(a) the black curves are plotted for $\Gamma = \Gamma_T = 0.0125$, 0.025 and 0.05. Other parameters are as in Figs. 5 and 6. To illustrate the influence of the SQUID on the tank circuit we include the grey curve with $\Gamma = \Gamma_T = 0.025$ and $\alpha = 0$. We see that the spectra exhibit spikes at multiples of the drive frequency $f_d = 0.1066$, with at least 6 harmonics clearly visible. The continuous part of each spectrum is highly structured, with a peak near $f = 0.09$ and a shoulder near $f = 0.02$. In the vicinity of the drive frequency the "background" of $s_{uT}$ (which we define as the average value of $s_{uT}$ some 1-3 channels away from the drive frequency) amounts to about 0.02 ($S_{UT} \approx 2k_BTR$) for $\Gamma = \Gamma_T = 0.05$, decreasing linearly with $\Gamma$. When $\alpha$ is set to zero, as shown by the grey curve for $\Gamma = 0.025$, $s_{uT}$ is greatly reduced for frequencies below about 0.3.

Figure 7(b) shows the low frequency power spectra $s_{vT}$, that is, the noise power of the Fourier component of $u_T$ at the drive frequency. The spectra are white, with average values of 0.0059, 0.012 and 0.028 for $\Gamma = \Gamma_T = 0.0125$, 0.025 and 0.05. Note that these values are roughly a factor of 3.5 lower than the values of the background of $s_{uT}$ near the drive frequency. One might have expected that the two values would coincide, at least roughly. However, $s_{vT}$ monitors only the amplitude fluctuations of $u_T$ and can thus be substantially lower than the background of $s_{uT}$ near $f_d$. Unfortunately, we found the ratio of $s_{uT}$ and $s_{vT}$ to depend on the various parameter values. An optimization of the noise energy using the background in $s_{uT}$ rather than $s_{vT}$, which would have been much faster than a direct optimization, is thus not possible. For the three $\Gamma$ values above we found transfer functions $dv_T/d\varphi_{ext}$ of -1.08, -1.09 and -1.12, respectively. We thus find $e = 0.52$, 0.52 and 0.56, respectively, with error bars of about ±10% that include scatter in both voltage noise power and transfer function.

To illustrate the low-frequency voltage noise power further, for the above parameters Fig. 8(a) shows the transfer function $dv_T/d\varphi_{ext}$, the voltage noise power $s_{vT}$ and the normalized noise energy $e$ as functions of $i_d$ for $\varphi_{ext} = 0.25$. We see that that $e$ is indeed lowest ($e \approx 0.5$) near the bias $i_d \approx 0.37$ found by the numerical optimization routine, although the transfer function is not optimized here. The voltage noise becomes quite large near the maxima of $dv_T/d\varphi_{ext}$, leading to an optimum bias current between these maxima. Figure 8(b) shows $dv_T/d\varphi_{ext}$, $s_{vT}$ and $e$ as functions of $\varphi_{ext}$ at $i_d = 0.369$. The minimum of $e$, about 0.45, is in fact near $\varphi_{ext} = 0.28$, that is, slightly above the value of 0.25 which had been fixed for optimization. To investigate the optimal parameter set further, we performed an optimization with variable applied flux, using as initial conditions the optimal parameters found for $\varphi_{ext} = 0.25$, $f_0 = 0.1$, $Q_0 = 100$ and $\beta_c = 0$. The result was a noise energy $e = 0.4$, with a transfer function of -1.12, at $\varphi_{ext} = 0.3$ and optimal parameters $f_d = 0.1066$, $i_d = 0.469$, $\alpha = 0.8$ and $\beta'_L = 1.09$. The reduction in $e$ compared to the value at $\varphi_{ext} = 0.25$ is only about 10% and we thus fix $\varphi_{ext} = 0.25$ for further calculations. We note here that a normalized noise energy $e \approx 0.4$ corresponds to a noise energy $\varepsilon \approx 0.8 k_B T \Phi_0 / I_0 R$ that is, $\varepsilon = 5 k_B T L / R$ for $\beta'_L = 1.09$. For comparison, a symmetric dc SQUID with an inductance parameter $\beta_L = 2 I_0 L / \Phi_0 = \beta'_L / \pi \approx 1$ has a noise energy $e \approx 2$ for $\Gamma = 0.05$, corresponding to $\varepsilon \approx 4 k_B T \Phi_0 / I_0 R \approx 8 k_B T L / R$. For the given values of $L/R$ and temperature the rf SQUID, with the parameters discussed here, has a low-frequency noise energy $\varepsilon$ about a factor of 8/5 lower than for this dc SQUID.

## 5. OPTIMIZED NOISE ENERGY

We next turn to the optimized noise energy. For the limit $\beta'_L \ll 1$ Chesca finds [18]

$$e_{BC,opt} = \frac{3}{4\pi}\left[(1+\beta'^2_L f^2)(1+\frac{\Gamma_T}{\Gamma}\frac{1}{\alpha^2 Q_0}\frac{1+\beta'^2_L f_d^2}{\beta'_L f_d})\right]\frac{\exp(\Gamma\beta'_L)}{\beta'_L} \quad . \tag{12}$$

This equation, which is for $\beta_c = 0$, is not yet optimized for $\alpha$ and $\beta'_L$. For $\beta'_L f_d \ll 1$ and $\Gamma\beta'_L \ll 1$ (12) reduces to $e_{BC,opt} \approx 0.24 \cdot (1+\Gamma_T/\Gamma\beta'_L\alpha^2 Q_0 f_d)/\beta'_L$, precisely the expression given by Danilov, et al. [17] For $\beta'_L = 1.21$, $Q_0 = 100$, $f_d = 0.1066$, $\alpha = 0.725$ and $\Gamma = \Gamma_T = 0.025$ we find $e_{BC,opt} \approx 0.23$, which is a factor of about 2 below the value obtained numerically. This difference is not unreasonable considering that the above values for $\beta'_L$ and $\alpha$ clearly stretch the limit of the analytic formula.

In terms of numerical calculations, Fig. 9 shows $e$ vs. $f_0$ as solid circles, with $f_0$ ranging from 0.01 to 1, for $Q_0 = 100$, $\varphi_{ext} = 0.25$ and $\beta_c = 0$. The noise energy was optimized with respect to $f_d$, $\alpha$, $\beta'_L$ and $i_d$. The graph also shows the values of $\alpha$ and $(dv_T/d\varphi_{ext})/30f_0$ (scaled to fit into the figure). The optimized parameter values, together with the resulting values of $dv_T/d\varphi_{ext}$, $s_{vT}$ and $e$, are also listed in Table 1. Throughout the frequency range investigated, $e$ stays almost constant, with a value near 0.5 (for comparison, from (12) we would have expected a quadratic increase in $e$). In all cases the optimal value of $\beta'_L$ is near 1, although the scatter for this and the other optimized parameters is appreciable, indicating that near optimum values for $e$ can be achieved over relatively large parameter ranges. However, at least for small drive frequencies, $\alpha$ should be quite large and is likely to be out of the experimental range. When, on the other hand, we fix $\alpha$ to the modest value of 0.2 in Fig. 10, we see that $e$ increases with decreasing $f_0$; at the lowest value shown, $f_0 = 1/60$, e $\approx 1.5$. As a guide to the eye we have plotted the function $0.4/f_0^{0.3}$, which approximately follows the computed points. In the graph we have also plotted the transfer function $|dv_T/d\varphi_{ext}|/30f_0$, which remains essentially constant at the value of 1; thus, when we include the factor $\gamma_L$, $dv_T/d\varphi_{ext} \approx -30\gamma_L f_0$.

How serious is the problem of undercoupling? Figure 11(a) displays $e$ vs. $\alpha$ for $f_0 = 0.5$, $\varphi_{ext} = 0.25$ and of $Q_0 = 50$, 100 and 200. One simulation for $Q_0 = 100$ was performed for $\Gamma = 0.025$, $\Gamma_T = 0$, all others for $\Gamma = \Gamma_T = 0.025$. The large value of $f_0$ was chosen to allow the calculations to be performed in a reasonable time. Figure 11(b) shows a similar plot for $f_0 = 0.1$ and $Q_0 = 100$ and 200. Selected values for the optimized parameters are listed in Table 2. For both values of $f_0$, $e$ increases with decreasing $\alpha$. For $f_0 = 0.5$ the curves for $Q_0 = 50$ and 100 can be described by the functional form [dotted lines in Fig. 11(a)] $0.5 + 0.33/Q_0\alpha^2$ suggested by (12), although the constants differ somewhat. This form works less well for $Q_0 = 200$, however, and for $f_0 = 0.1$ this dependence also does not fit the data. In Fig. 11(b) we have plotted the empirical function $0.38/\alpha^{0.6}$ (dashed line) that fits the data reasonably well for $Q_0 = 100$. For $f_0 = 0.1$ and $\alpha = 0.2$ we also investigated the $Q_0$ dependence of $e$ in more detail and found that, for $Q_0$ between 50 and 400, $e$ decreased by only about 25%, that is, the dependence on $Q_0$ is very weak, indicating that the effective quality factor of the system which we define via $v_T = Q_{eff}\beta'_L f_0 i_d$ (in absolute units: $V_T = Q_{eff} L\omega_0 I_d$) is dominated by the SQUID. For example, for $Q_0 = 200$, $\alpha = 0.2$ and $i_d = 0.15$ listed in Table 2, we find a tank circuit voltage $v_T = 0.88$, yielding $Q_{eff} \approx 55$, much less than $Q_0$. Another interesting prediction of (12) is that, for $\Gamma_T = 0$, $e$ should be independent of $\alpha$ and $Q_0$. As can be seen in Fig. 11(a) for $f_0 = 0.5$ and $Q_0 = 100$ we indeed observe that $e$ is independent of $\alpha$ for $0.1 \lesssim \alpha \lesssim 1$; however, $e$ increases below $\alpha \approx 0.1$. Furthermore, Fig. 11(b) shows that for $f_0 = 0.1$ and $Q_0 = 100$ $e$ is strongly reduced for $\Gamma_T = 0$ and $\alpha \lesssim 0.3$. In Fig. 11(b) we have plotted the fitting function $0.45[(1-\alpha)^2 + 1]$ which approximately fits the data. We again emphasize, however, that, due to the error margins of the numerical data, such fit functions should not be taken too literally.

We next turn to the dependence of $e$ on $\beta'_L$ and $\beta_c$. As we have already seen, the lowest values were found for $\beta'_L$ slightly above unity. Figure 12(a) shows simulations for $f_0 = 0.5$, $\Gamma = \Gamma_T = 0.025$, $Q_0 = 100$ and $\varphi_{ext} = 0.25$ for $\beta_c = 0$ and 1; Fig. 12(b) shows the results for $f_0 = 0.1$. As we see, in all cases there is indeed a pronounced minimum in $e$ ($\approx 0.5$) for $\beta'_L$ slightly above 1. Optimized values for $\alpha$ were in the range 0.25 – 0.6 for $f_0 = 0.5$ and above 0.8 for $f_0 = 0.1$ (cf. Table 3). We further see that there is no significant difference for the two $\beta_c$ values nor for the two drive frequencies as long as $\beta'_L$ is below about 1.5; at higher values of $\beta'_L$, $\beta_c = 0$ yields much better results. We also compared the numerical curves in Fig. 12 with the functional dependences suggested by (12) using $e \approx 0.4 \cdot (1 + f_0^2 \beta'^2_L)/\beta'_L$ (that is, we ignore the contribution from the tank circuit; we also increased the prefactor $3/4\pi \approx 0.24$ to 0.4). The agreement of the scaled analytical formula is excellent up to $\beta'_L \approx 1$, where a linear increase of $e$ takes over, fitted approximately in Fig. 12(a) by $e \approx 0.36 \beta'_L$. For large values of $\beta'_L$, the intrinsic noise energy is predicted to be $\varepsilon \approx I_0^2 L \Gamma^{4/3}/2\omega_{rf}$, or in our notation, $e \approx \beta'_L \Gamma^{1/3}/8\pi f_d$. The latter expression displays the scaling with $\beta'_L$. The prefactor, with $f_d \approx 0.5$ and $\Gamma = 0.025$, amounts to only 0.023 and is thus much lower than the value 0.36 used for the fit. However, the analytical formula is valid in the limit of small drive frequencies which is not fulfilled here.

Finally, we examine the issue of $\beta_c$, which appears in several ways. First, it determines the damping of the isolated junction. In a quasistatic situation (that is, for low enough drive frequencies) and for large values of $\beta'_L$, when multiple transitions in the quantum states of the SQUID loop are possible, the switching between these states becomes irregular when $\beta_c$ is large, leading to increased noise. This effect may explain why $e$ increases so strongly with $\beta_c$ for $\beta'_L \gtrsim 2$. For situations where $\beta'_L < 1$ or when only two flux quantum states are possible the switching effect should be greatly reduced, and this is indeed observed in Fig. 12 where the curves $\beta_c = 0$ and $\beta_c = 1$ essentially coincide. To study dynamic effects, we should investigate two more characteristic frequencies: the junction plasma frequency, $f_{pl} = 1/\sqrt{\beta_c}$ in normalized units, and the (normalized) $LC$ resonance frequency of the SQUID loop, $f_{LC} = 1/2\pi f_c / \sqrt{LC} = 1/\sqrt{\beta'_L \beta_c}$. For $\beta'_L$ values near or below the optimum of $e$ vs. $\beta'_L$, $f_{LC} \gtrsim f_{pl}$, while for $\beta_c \rightarrow \infty$ both frequencies tend to zero. More realistically, both the plasma and $LC$ resonance frequencies are on the order of several gigahertz. Practical drive frequencies are thus in general below resonance, in which case we expect that $e$ becomes more or less independent of $\beta_c$. When, on the other hand, the drive frequency approaches $f_{LC}$ or $f_{pl}$ we expect chaotic effects (and perhaps even stable parameter ranges where the noise performance becomes better than that discussed in this paper). Furthermore, when $f_d$ becomes larger than $f_{LC}$ the SQUID loop can no longer follow the oscillating flux drive, and one expects the device to cease to function (indeed, we have seen this in simulations). To examine the effect of $\beta_c$ more closely we choose $\beta'_L = 1.5$ and take the parameters of Fig. 12(b) as starting conditions to vary $\beta_c$. Setting $i_d = 0.49$, $f_d = 0.11$ ($\xi = 22.75$), $\alpha = 0.49$, $Q_0 = 100$, $f_0 = 0.1$, $\varphi_{ext} = 0.25$ and $\Gamma = \Gamma_T = 0.025$, and increasing only $\beta_c$ we find the modulus of the transfer function decreases (Fig. 13, open circles) and the noise energy $e$ increases, reaching a value of about 1 for $\beta_c = 10$ (Fig. 13, open squares). On the other hand, when we vary $i_d$ and $f_d$, the decrease in $|dv_T/d\varphi_{ext}|$ is more modest (Fig. 13, solid circles), and $e$ remains low for $\beta_c < 35$ (Fig. 13, solid squares). For larger values of $\beta_c$, $e$ increases strongly. At $\beta_c = 35$ the SQUID loop $LC$ frequency is 0.14, comparable to the drive frequency $f_d = 0.11$.

Thus, we see that in the parameter range of interest $\beta_c$ can be allowed to reach substantial values. Consequently, only a moderate damping of the junction is required and $I_0R$ can be much larger than for dc SQUIDs where $\beta_c < 1$ is mandatory. Thus, even for comparable values of $e$ the dimensioned noise energy $\varepsilon$ can be lower for rf SQUIDs than for dc SQUIDs. Is shunting necessary at all? We believe so, because otherwise the nonlinear quasiparticle resistance, of, say, an Nb tunnel junction would lead to additional up and down conversion of noise and probably also to additional chaotic effects.

## 6. CONCLUSIONS

We have seen from our numerical analysis that SQUID noise energies $e$ of 0.4-0.5 can be achieved over a wide parameter range. The optimal value for $\beta'_L$ is 1 or slightly higher. In absolute units we thus obtain $\varepsilon \approx (0.8-1)k_BT\Phi_0/I_0R$ or $\varepsilon \approx (5-6)k_BTL/R$. In the latter formulation the noise energy for the rf SQUID is not very different from that for the dc SQUID [$\varepsilon \approx (8-9)k_BTL/R$]. However, for the dc SQUID $\beta_L = 2I_0L/\Phi_0 = \beta'_L/\pi$ should be about 1. Thus, for a given value of $L$, the rf SQUID allows for a lower value of $I_0$ than does the dc SQUID. When we fix $I_0R$ for a given junction, ε is a factor of 4-5 lower for the rf SQUID than for the dc SQUID. Furthermore, for the dc SQUID $\beta_c$ must be lower than 1 while for the rf SQUID it can be substantially higher in the regime where $\beta'_L$ is near optimum. If $\beta_c$ is determined by shunting the junction, $R$ and thus $I_0R$ for the rf SQUID can be by much higher than for the dc SQUID. The above discussion was for the low temperature limit; however, we also saw that the rf SQUID tolerates much higher values of the noise parameter $\Gamma$ – which can be about unity – without degrading the noise performance significantly.

What are the drawbacks of the rf SQUID? We saw that undercoupling is a problem even for drive frequencies as high as 10% of the junction characteristic frequency (that is, $f_0 = 0.1$). Using $f_c = I_0 R/\Phi_0 = 100$ GHz as a typical value for the junction characteristic frequency, realistic drive frequencies are not much greater than 0.01. As indicated by our frequency dependent calculations (Fig. 10) normalized noise energies of 1.5-2 – which are comparable to the case of dc SQUIDs – should still be feasible, provided that the coupling parameter $\alpha$ can be made as large as 0.2. We have further seen that, in contrast to the analytical predictions for low values of $\alpha$ and $\beta'_L$, for $\beta'_L$ of order unity a scaling of $e$ as $1/\alpha^2 Q_0$ is not observed. Increasing $Q_0$ well beyond 100 leads to a saturation of $e$, indicating that the effective quality factor is dominated by dissipation in the SQUID. It thus seems that the focus should be on making $\alpha$ as large as possible rather than choosing very large values of $Q_0$ at the expense of small values of $\alpha$.

We now give a realistic example, assuming that $\alpha = 0.2$ is feasible. Figure 14(a) shows $i_d$ vs. $v_T$ together with $v_T$ vs. $\varphi_{ext}$ (upper inset) and $dv_T/d\varphi_{ext}$ vs. $i_d$ (lower inset) for a device with $Q_0 = 100$, $f_0 = 0.1$, $f_d = 0.1011$ and $\beta'_L = \beta_c = 1$. In Fig. 14(b), we have plotted $e$, $|dv_T/d\varphi_{ext}|$ and $s_{vT}/\Gamma$ vs. $\Gamma$. We have fitted the noise energy with the expression $0.95 \exp\Gamma$. Assuming the SQUID parameters $f_c = 100$ GHz and $L = 50$ pH and $T = 4.2$ K, we find $I_0 R \approx 207$ µV, and with $\beta'_L = 1$ we obtain $I_0 \approx 6.6$ µA, $R \approx 31$ Ω and $\Gamma \approx 0.027$. We note that $R$ is rather large but should easily be accessible experimentally. For $\alpha = 0.2$ and $f_0 = 0.1$ (10 GHz), $e \approx 0.95$ yielding a noise energy $\varepsilon \approx 10\hbar$. For $\alpha = 0.2$ and $f_0 = 0.01$, using the dependence of $e$ vs. $f_0$ shown in Fig. 10, we see that the noise energy would be approximately doubled. Let us further assume that for a 10 GHz (1 GHz) drive we have $L_T = 5$ nH (50 nH). The reduced low frequency voltage noise power $s_{vT}/\Gamma \approx 5.5$ at 10 GHz then corresponds to $S_V^{1/2} \approx 1$ nV/Hz$^{1/2}$; for the 1-GHz drive $S_V^{1/2}$ is reduced by about a factor of 3. For the two drive frequencies $C_T$ should be about 50 fF (0.5 pF) and, with $Q_0 = 100$, we obtain $R_T \approx 3\Omega$ in both cases. These numbers are certainly attainable experimentally and a device like this should be feasible.

We briefly comment on some of the many remaining open questions. One question regards the detuning parameter $\xi$. Most of our optimizations yielded positive values for $\xi$, that is, the device was driven slightly above resonance. We performed some simulations with negative $\xi$ and also found noise energies of about 0.5. However, in general, negative values of $\xi$ seemed to result in more instabilities (jumps and hystereses in the $v_T$ vs. $\varphi_{ext}$ characteristics) and were thus less accessible to our optimization procedures. Another question regards the regime of large values of $\beta'_L$. An important parameter introduced in previous analyses of the dissipative regime [4-8] was the slope parameter $\eta$ describing the finite slope of the current steps in the $i_d$ vs. $v_T$ curves (an extrinsic noise term proportional to $\eta$ actually dominates the rf SQUID performance in the dissipative regime). It would be interesting to evaluate such dependencies and the noise performance in parameter ranges far from optimum. Unfortunately, systematic simulations on these issues are much too time-consuming to be included in the present investigation. Furthermore, one can distinguish a variety of different regimes of operation, for example as classified by the various characteristic (resonance) frequencies of the junction, the SQUID loop and the tank circuit and their relative ratios. A thorough discussion is again far beyond the scope of this paper.

Finally, another quantity of considerable interest for SQUID amplifiers – the noise temperature – is analyzed in the paper that follows. The noise temperature requires the analysis of a more complete circuit, including an input circuit coupled to the SQUID loop and an output preamplifier. We shall see that the optimal noise temperature can reach very low values even for realistic parameters, hopefully stimulating further experimental efforts to improve the noise properties of these devices.

**ACKNOWLEGMENTS**

The authors thank A. I. Braginski, B. Chesca, M. Mück and Y. Zhang for valuable discussions. Financial support by the Deutsche Forschungsgemeinschaft (R. K. and D. K) is gratefully acknowledged. This work was also supported by the Director, Office of Science, Office of Basic Energy Sciences, Materials Sciences and Engineering Division, of the U.S. Department of Energy under Contract No. DE-AC02-05CH11231 (JC).

# APPENDIX A: LIST OF SYMBOLS

$C$: junction capacitance

$C_T$: capacitance of tank circuit

$e = \pi s_\varphi / 2\Gamma \beta'_L = \varepsilon / [2\Phi_0 k_B T / I_0 R]$: normalized low frequency noise energy

$f_c = I_0 R / \Phi_0$: junction characteristic frequency

$f_d$: drive frequency (used either in absolute units or normalized to $f_c$)

$f_0 = \Phi_0 / 2\pi I_0 R \sqrt{L_T C_T} = 1/2\pi f_c \sqrt{L_T C_T}$: normalized unloaded tank circuit resonance frequency

$f_{LC} = 1/2\pi f_c / \sqrt{LC} = 1/\sqrt{\beta'_L \beta_c}$: normalized LC resonance of SQUID loop

$f_{pl} = 1/\sqrt{\beta_c}$: normalized junction plasma frequency

$i_T = I_T/I_0$: normalized bias current of tank circuit

$\tilde{i}_T = v_T / \beta'_L \gamma_L^2 f$

$i_d = I_d/I_0$: normalized amplitude of drive current of tank circuit

$i_N = \sqrt{2\Gamma / \tau_{noise}} \cdot i_{N,0}$ normalized noise current in SQUID

$i_{N,0}$: Gaussian distributed random numbers with zero average and mean square deviation of 1

$i_{NT,0}$: Gaussian distributed random numbers with zero average and mean square deviation of 1

$I_0$: junction critical current in the absence of noise

$I_d$: amplitude of oscillating drive current

$I_T$: bias current of tank circuit

$I_1, I_2$: currents through arms of tank circuit

$I_N$: noise current in SQUID loop

$I_{N,T}$: noise current in tank circuit

$j = J/I_0$: normalized circulating current in SQUID

$J$: circulating current in SQUID loop

$k_B$: Boltzmann constant

$L$: SQUID inductance

$L_T$: inductance of tank circuit

$M$: mutual inductance between SQUID and tank circuit

$Q$: quality factor (general)

$Q_0 = \sqrt{L_T / C_T} / R_T$: unloaded quality factor of tank circuit

$Q_{eff} = v_T / \beta'_L f_0 i_d = V_T / L\omega_0 I_d$: effective quality factor

$R$: junction resistance

$R_T$: tank circuit resistance

$S_I = 4k_BT/R$: spectral density of current noise power in SQUID loop

$S_{IT} = 4k_BT_T/R_T$: spectral density of current noise power in tank circuit

$S_{UT}$: spectral density of high frequency voltage noise across tank circuit

$S_{VT}$: spectral density of low frequency voltage noise across tank circuit

$s_{vT} = S_{V,T}/[I_0R\Phi_0/2\pi]$: spectral density of normalized low frequency voltage noise

$s_{vT} = S_{U,T}/[I_0R\Phi_0/2\pi]$: spectral density of normalized high frequency voltage noise

$s_\varphi = s_{vT}/(dv_T/d\varphi_{ext})^2$: spectral density of normalized low frequency flux noise

$T$: temperature of SQUID

$T_T$: temperature of tank circuit

$u = U/I_0R$: normalized voltage

$u_{NT} = \gamma_R \cdot i_{NT} = \sqrt{\beta'_L f_0 \gamma_L^2 / Q_0} \cdot \sqrt{2\Gamma_T / \tau_{noise}} \cdot i_{NT,0}$: normalized noise voltage across tank circuit resistor

$U_{CT}$: voltage across tank circuit capacitor

$U_{LT}$: voltage across tank circuit inductor

$U_{NT}$: noise voltage across tank circuit resistor

$U_{RT}$: voltage across tank circuit resistor

$v_T$: normalized voltage amplitude across tank circuit at drive frequency

$\alpha = M/\sqrt{LL_T}$: coupling parameter

$\beta_c = 2\pi I_0 R^2 C / \Phi_0$: Stewart-McCumber parameter

$\beta'_L = 2\pi I_0 L / \Phi_0$: inductance parameter

$\gamma_L = \sqrt{L_T/L}$: inductance scaling parameter; throughout the manuscript calculations are for $\gamma_L = 1$;

   scaling with $\gamma_L$:

   $u_{cT}(\gamma_L) = u_{cT}(\gamma_L = 1) \cdot \gamma_L$, $\varphi_T(\gamma_L) = \varphi_T(\gamma_L = 1) \cdot \gamma_L$, $i_d(\gamma_L) = i_d(\gamma_L = 1)/\gamma_L$;

   $v_T(\gamma_L) = v_T(\gamma_L = 1) \cdot \gamma_L$; $dv_T/d\varphi_{ext}(\gamma_L) = dv_T/d\varphi_{ext}(\gamma_L = 1) \cdot \gamma_L$;

   $s_v^{1/2}(\gamma_L) = s_v^{1/2}(\gamma_L = 1) \cdot \gamma_L$; $s_\varphi(\gamma_L) = s_\varphi(\gamma_L = 1)$; $\varepsilon(\gamma_L) = \varepsilon(\gamma_L = 1)$

$\gamma_R = R_T/R = \beta'_L f_0 \gamma_L^2 / Q_0$: ratio of tank circuit resistor to junction resistor

$\Gamma = 2\pi k_B T / I_0 \Phi_0$: noise parameter for SQUID

$\Gamma_T = 2\pi k_B T_{tank} / I_0 \Phi_0$: noise parameter for tank circuit

$\delta$: gauge invariant phase difference across Josephson junction

$\varepsilon$: low frequency noise energy

$\varphi = \Phi/\Phi_0$: normalized flux through SQUID loop

$\varphi_T = 2\pi\Phi_T/\Phi_0$: normalized flux through tank circuit inductor

$\Phi$: total flux through SQUID loop

$\Phi_{ext}$: applied flux

$\Phi_T$: flux through tank circuit inductor

$\Phi_0$: flux quantum

$\tau = \Phi_0 / 2\pi I_0 R$: normalized time

$\tau_{noise}$: normalized time step for noise calculations

$\xi = 2Q_0(f_d/f_0 - 1)$: detuning parameter

# References

1. C.D. Tesche and J. Clarke, *J. Low Temp. Phys.* **29**, 301 (1977).

2. J.J.P. Bruines, V.J. de Waal, J.E. Mooij, *J. Low Temp. Phys.* **46**, 383 (1982).

3. For a recent overview, see, for example, B. Chesca, R. Kleiner and D. Koelle, in: *The SQUID Handbook, Vol. 1*, J. Clarke & A. I. Braginski (eds), Wiley-VCH, Weinheim (2004) p. 29.

4. J. Kurkijärvi, *Phys. Rev. B* **6**, 832 (1972).

5. J. Kurkijärvi, *J. Appl. Phys.* **44**, 3729 (1973).

6. J. Kurkijärvi and W.W. Webb, in: *Proc. Appl. Supercond.* IEEE Pub. No. 72CHO 682-5-TABSC (IEEE, New York, 1972) p. 581.

7. L.D. Jackel, and R.A. Buhrman, *J. Low Temp. Phys.* **19**, 201 (1975).

8. G.J. Ehnholm, *J. Low Temp. Phys.* **29**, 1 (1977).

9. R.P. Giffard and J.N. Hollenhorst, *Appl. Phys. Lett.* **32**, 767 (1978).

10. J.N. Hollenhorst and R.P. Giffard, *J. Appl. Phys.* **51**, 1719 (1980).

11. P.K. Hansma, *J. Appl. Phys.* **44**, 4191 (1973).

12. P.K. Hansma, *Phys. Rev. B* **12**, 1707 (1975).

13. R.Rifkin, D.A. Vincent, B.S. Deaver, Jr., and P.K. Hansma, *J. Appl. Phys.* **47**, 2645 (1976).

14. O.H. Sorensen, *J. Appl. Phys.* **47**, 5030 (1976).

15. K.K. Likharev, B.T. Ulrich, *Dynamics of Josephson junctions Circuits: Basic Theory*, Moscow Univ. Publ., Moscow (1978) [in Russian]; see also K. K. Likharev, *Dynamics of Josephson Junctions and Circuits*, Gordon and Breach Science Publishers, Amsterdam (1986).

16. V.V. Danilov and K.K. Likharev, *Zh. Tekhn. Fiz.* **45**, 1110 (1975) [*Sov. Phys. Techn. Phys.* **20**, 697 (1976)].

17. V.V. Danilov, K.K. Likharev, and O.V. Snigirev, *Proceedings of SQUID '80*, Walter de Gruyter & Co., Berlin (1980) p. 473.

18. B. Chesca, *J. Low Temp. Phys.* **110**, 963 (1998).

19. R. Kleiner, D. Koelle, and John Clarke (2006), "A Numerical Treatment of the Rf SQUID: II. Noise Temperature", *J. Low Temp. Phys.* (submitted).

20. W.C. Stewart *Appl. Phys. Lett.* **12**, 277 (1968).

21. D.E. Mc Cumber *J. Appl. Phys.* **39**, 3113 (1968).

22. D. Koelle, R. Kleiner, F. Ludwig, E. Dantsker, and John Clarke, *Rev. Mod. Phys.* **71**, 631 (1999).

**Figure captions**

Fig. 1. The rf SQUID inductively coupled to its tank circuit.

Fig. 2. Example of the evolution of $e$ and the parameters $\beta'_L$, $\alpha$, $i_d$, $f_d$ (with initial values shown inset) during the optimization procedure. Fixed parameters were $Q_0 = 100$, $\varphi_{ext} = 0.25$, $f_0 = 0.1$, $\beta_c = 0$ and $\Gamma = \Gamma_T = 0.025$. In the graph the detuning parameter $\xi = 2Q_0(f_d/f_0 - 1)$ rather than $f_d$ is plotted.

Fig. 3. Zero temperature characteristics for the rf SQUID. (a) $i_d$ vs. $v_T$; (b) transfer function vs. $i_d$ at $\varphi_{ext} = 0.25$; and (c) $v_T$ vs. $\varphi_{ext}$ at $i_d = 0.8$. Solid circles are numerical calculations, solid lines are after Hansma [Eq. (10)] and dotted lines are after Chesca [Eq. (11)]. Parameters are $Q_0 = 100$, $f_0 = 0.1$, $f_d = 0.101$, $\beta'_L = 0.5$, $\beta_c = 0.5$ and $\alpha = 0.1$.

Fig. 4. Zero temperature characteristics for the rf SQUID. (a) $i_d$ vs. $v_T$; (b) transfer function vs. $i_d$ at $\varphi_{ext} = 0.25$; and (c) $v_T$ vs. $\varphi_{ext}$ at $i_d = 0.8$. Solid circles are numerical calculations, solid lines are after Hansma [Eq. (10)] and dotted lines are after Chesca [Eq. (11)]. Parameters are $Q_0 = 100$, $f_0 = 0.1$, $f_d = 0.101$, $\beta'_L = 0.5$, $\beta_c = 0.5$ and $\alpha = 0.3$.

Fig. 5. Characteristics for the rf SQUID for $\Gamma = \Gamma_T = 0.025$ (solid circles). (a) $i_d$ vs. $v_T$; (b) transfer function vs. $i_d$ at $\varphi_{ext} = 0.25$, and (c) $v_T$ vs. $\varphi_{ext}$ at $i_d = 0.369$. Parameters are $Q_0 = 100$, $f_0 = 0.1$, $f_d = 0.1066$, $\beta'_L = 1.21$, $\beta_c = 0$ and $\alpha = 0.725$. Open circles in (a) correspond to calculation for $\Gamma = 0$, dashed line shows Hansma's result, Eq. (10), for comparison. Arrow in (b) indicates the bias point for the calculation of curve (c). Tilted arrows in (c) indicate sweep direction, vertical arrows voltage jumps.

Fig. 6. Numerically calculated transfer functions. (a) $dv_T/d\varphi_{ext}$ vs. $i_d$ for 4 values of $\Gamma = \Gamma_T$, and (b) transfer function at $i_d = 0.369$ vs. $\Gamma$ for $\Gamma_T = \Gamma$ (black circles), $\Gamma_T = 0$ (open circles) and $\Gamma_T = 10\Gamma$ (grey circles). Other parameters are indicated in the figure.

Fig. 7. Spectral densities of noise power. (a) High frequency voltage $u_T$ across the inductor $L_T$ and (b) the low-frequency voltage noise $v_T$ across the tank circuit for three values of $\Gamma = \Gamma_T$. Parameters are listed in (a). In (a) a curve for $\alpha = 0$ and $\Gamma = \Gamma_T = 0.025$ has been included (grey line). The spectra have been averaged 100 times in (a) and 20 times in (b).

Fig. 8. Normalized low frequency voltage noise spectral density $s_{vT}$ (grey circles), normalized noise energy $e$ (black circles) and transfer function $dv_T/d\varphi_{ext}$ (open circles) as a function of (a) $i_d$, (b) $\varphi_{ext}$ and (c) $\Gamma = \Gamma_T$ for the same parameters as in Fig. 5(b). In (a) and (b), $s_v$ has been multiplied by 10. Arrow in (a) indicates optimal bias.

Fig. 9. Normalized noise energy (solid circles), optimized coupling constant $\alpha$ (open circles) and modulus of the optimized transfer function divided by $30f_0$ (open squares) vs. tank circuit resonance frequency $f_0$. Fixed parameters are $Q_0 = 100$, $\beta_c = 0$, $\varphi_{ext} = 0.25$ and $\Gamma = \Gamma_T = 0.025$. Optimized parameters are $\alpha, \beta'_L, i_d$ and $f_d$.

Fig. 10 Normalized noise energy $e$ and transfer function, re-scaled as $|dv_T/d\varphi_{ext}|/30f_0$, vs. $f_0$. Fixed parameters were $\alpha = 0.2$, $\varphi_{ext} = 0.25$, $\beta_c = 0$, $Q_0 = 100$ and $\Gamma = \Gamma_T = 0.025$. Parameters $\beta'_L, i_d$ and $f_d$ were varied to optimize $e$.

Fig. 11. Normalized noise energy $e$ vs. $\alpha$ for several values of $Q_0$. (a) $f_0 = 0.5$, (b) $f_0 = 0.1$. Other fixed parameters are $\varphi_{ext} = 0.25$, $\beta_c = 0$ and $\Gamma = 0.025$. In both graphs one curve is for

$\Gamma_T = 0$ while the others are for $\Gamma = \Gamma_T$. Parameters $\beta'_L$, $i_d$ and $f_d$ have been varied to optimize $e$. Dashed and dotted lines correspond to the analytical functions indicated in the figures. Some values of the optimized parameters are given in Table 2.

Fig. 12. Normalized noise energy $e$ vs. $\beta'_L$ for two values of $\beta_c$. (a) $f_0 = 0.5$, (b) $f_0 = 0.1$. Other fixed parameters are $Q_0 = 100$, $\varphi_{ext} = 0.25$ and $\Gamma = 0.025$. Parameters $\alpha$, $i_d$ and $f_d$ have been varied to optimize $e$. Dashed and dotted lines correspond to the analytical functions indicated in the figure. Some values of the optimized parameters are given in Table 3.

Fig. 13. Noise energy and modulus of transfer function vs. $\beta_c$. Circles show $|dv_T/d\varphi_{ext}|$ and squares show $e$. Open symbols are for fixed parameters $\beta'_L = 1.5$, $i_d = 0.49$, $f_d = 0.11$, $\alpha = 0.49$, $Q_0 = 100$, $f_0 = 0.1$, $\varphi_{ext} = 0.25$ and $\Gamma = \Gamma_T = 0.025$. Solid symbols indicate that $i_d$ and $f_d$ have been optimized.

Fig. 14. Example of numerical results for an rf SQUID with $\alpha = 0.2$. Parameters are $Q_0 = 100$, $f_0 = 0.1$, $f_d = 0.1011$, $\beta'_L = 1$ and $\beta_c = 1$. For $\Gamma = \Gamma_T = 0$ (a) shows $i_d$ vs. $v_T$ for $\varphi_{ext} = 0$, 0.25 and 0.5, $v_T$ vs. $\varphi_{ext}$ for $i_d = 0.24$ (upper inset) and $dv_T/d\varphi_{ext}$ vs. $i_d$ for $\varphi_{ext} = 0.25$ (lower inset). (b) Noise energy, the fitting function $e = 0.95 \exp(\Gamma)$, modulus of transfer function and $s_{vT}/\Gamma$ vs. $\Gamma$.

## Table captions

Table 1. Selected parameter values and some resulting quantities for the graphs in Fig. 9 ($e$ vs. $f_0$). Fixed parameters are $Q_0 = 100$, $\varphi_{ext} = 0.25$ and $\beta_c = 0$; fixed values of $f_0$ are listed. Optimized parameters: $i_d$, $\xi = 2Q_0(f_d/f_0-1)$, $\alpha$, and $\beta'_L$. Resulting quantities: normalized transfer function $dv_T/d\varphi_{ext}$, normalized voltage noise spectral density $s_{vT}$ and normalized noise energy $e$.

Table 2. Selected parameter values and some resulting quantities for the graphs in Fig. 11 ($e$ vs. $\alpha$). Fixed parameters are $\varphi_{ext} = 0.25$ and $\beta_c = 0$; fixed values of $\alpha$, $f_0$, $Q_0$ and $\Gamma_T$ are listed. Optimized parameters: $i_d$, $\xi$ and $\beta'_L$. Resulting quantities: normalized transfer function $dv_T/d\varphi_{ext}$ and normalized noise energy $e$.

Table 3. Selected parameter values and some resulting quantities for the graphs in Fig. 12 ($e$ vs. $\beta'_L$). Fixed parameters are $Q_0 = 100$, $\varphi_{ext} = 0.25$ and $\Gamma = \Gamma_T = 0.025$; fixed values of $\beta'_L, \beta_c$ and $f_0$ are listed. Optimized parameters: $i_d$, $\xi$ and $\alpha$. Resulting quantities: normalized transfer function $dv_T/d\varphi_{ext}$ and normalized noise energy $e$.

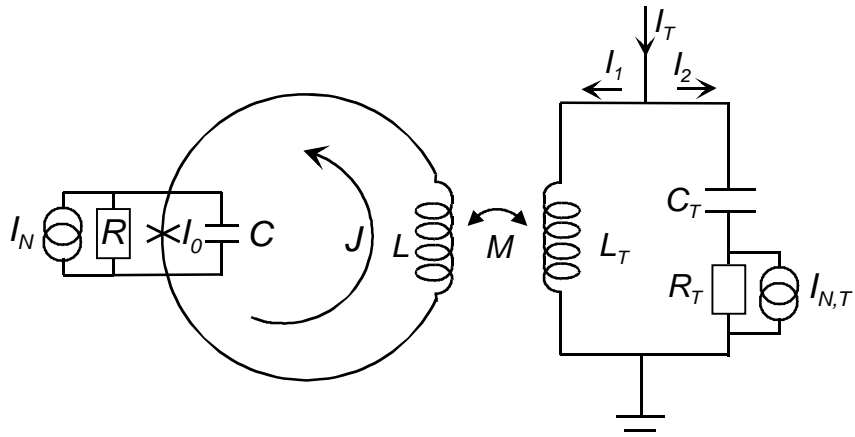

**Figure 1**

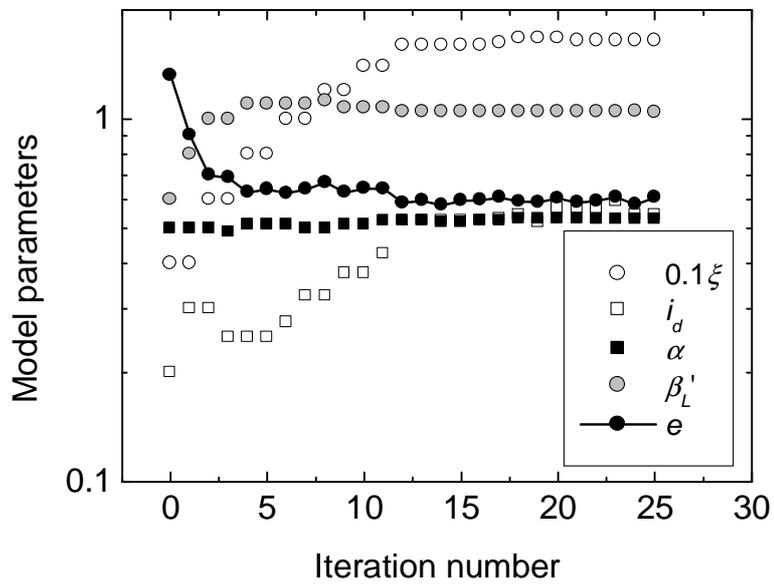

**Figure 2**

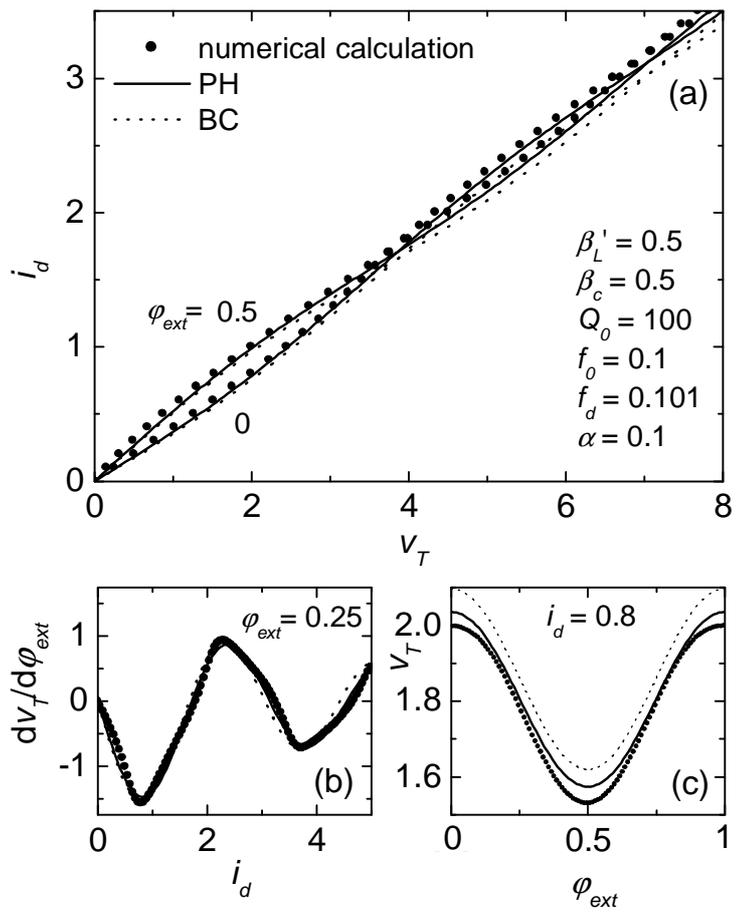

**Figure 3**

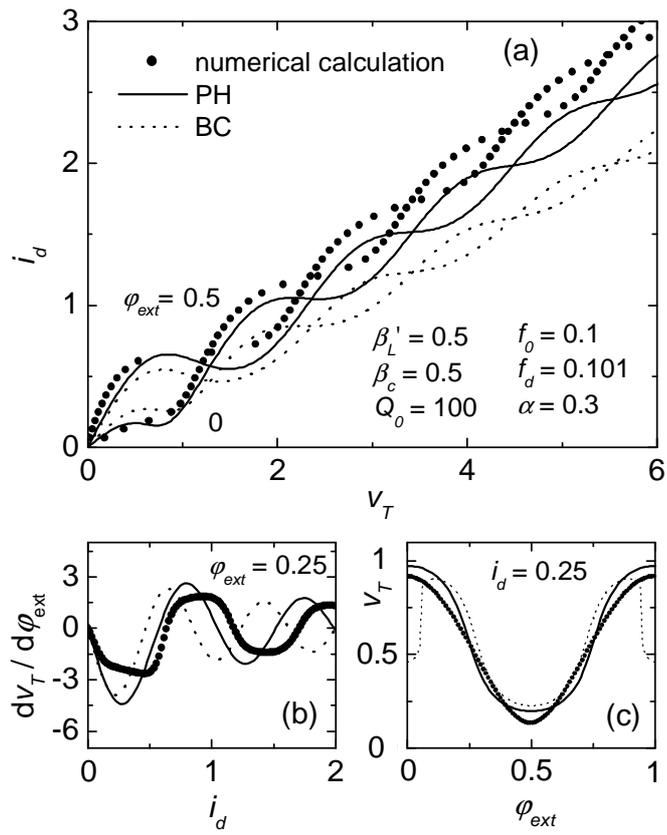

**Figure 4**

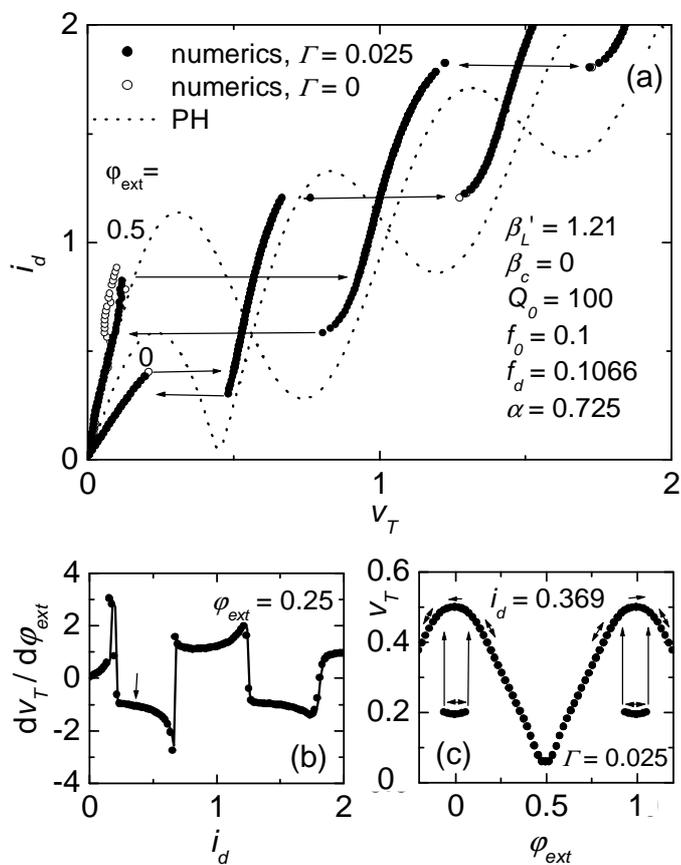

**Figure 5**

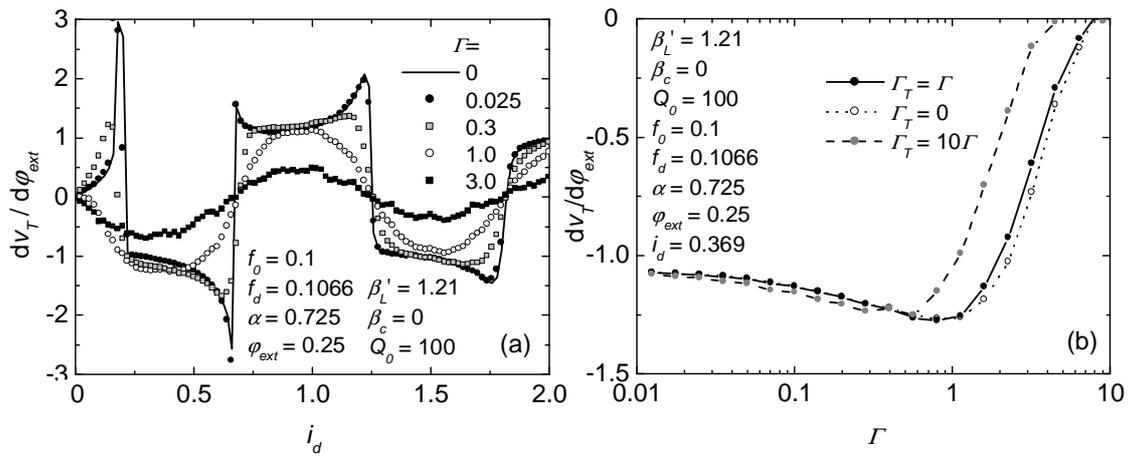

**Figure 6**

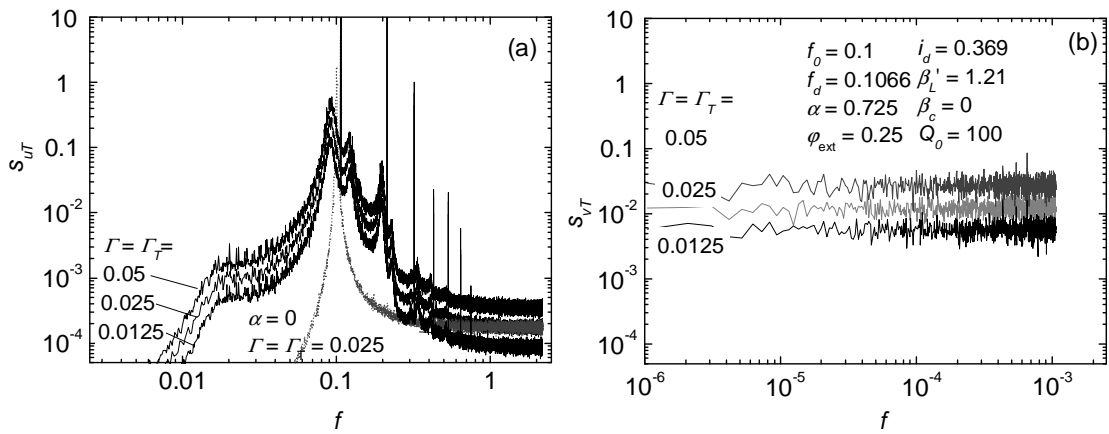

**Figure 7**

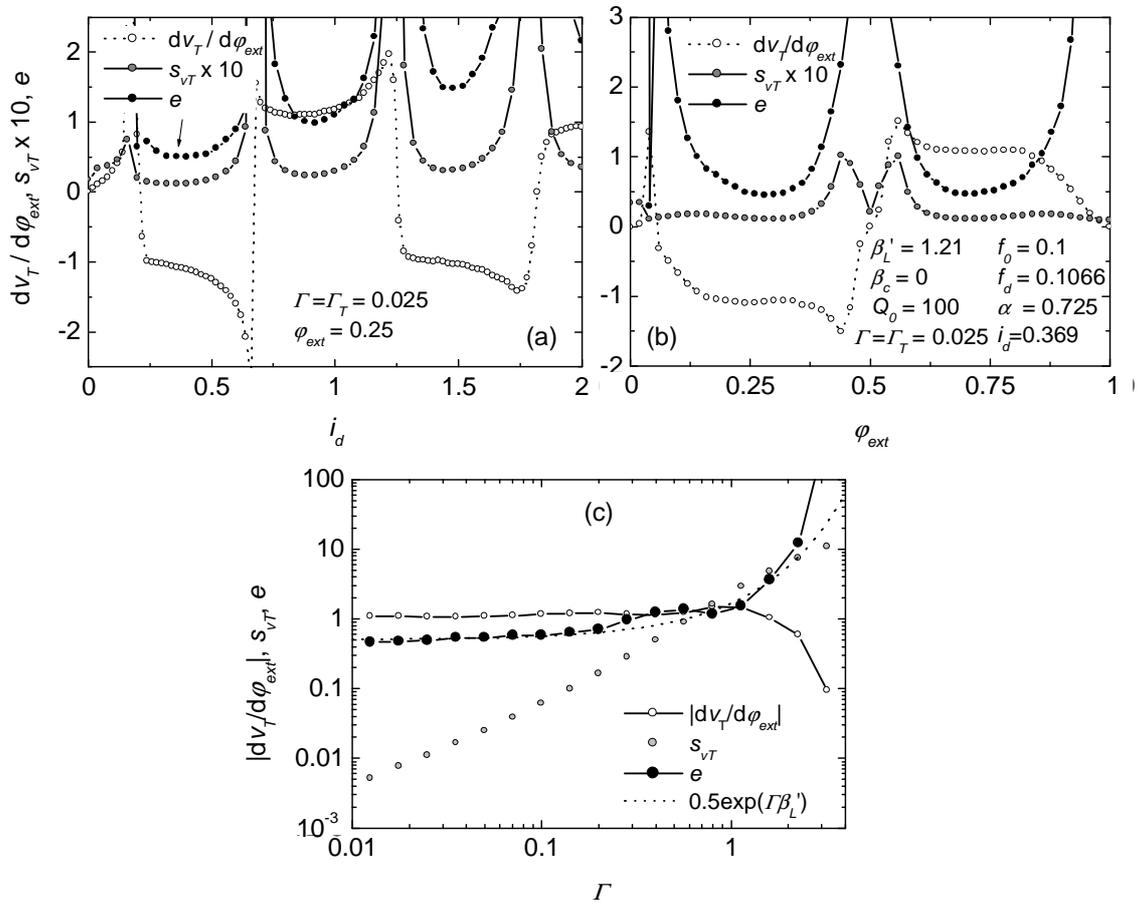

**Figure 8**

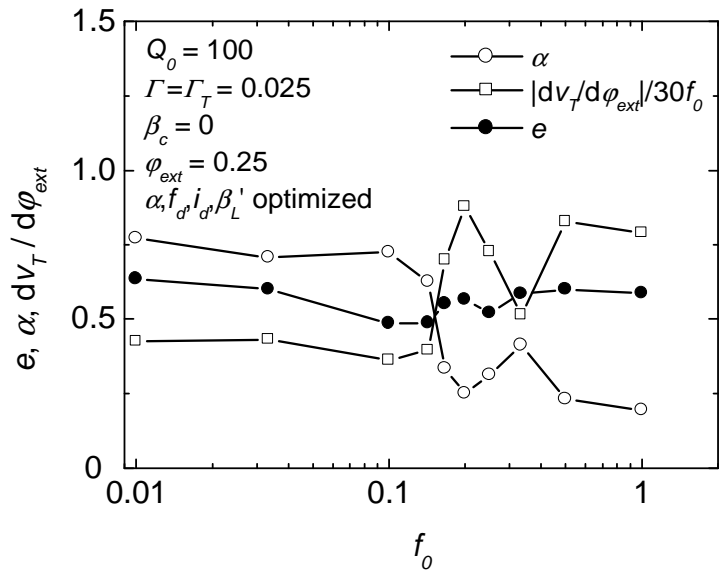

**Figure 9**

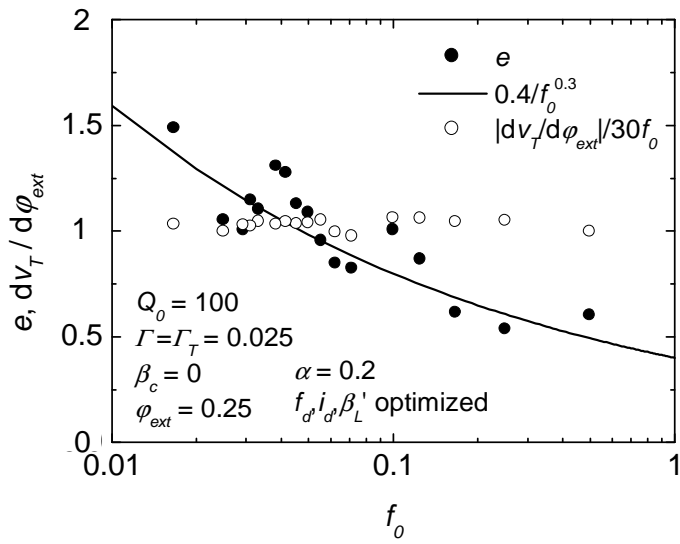

**Figure 10**

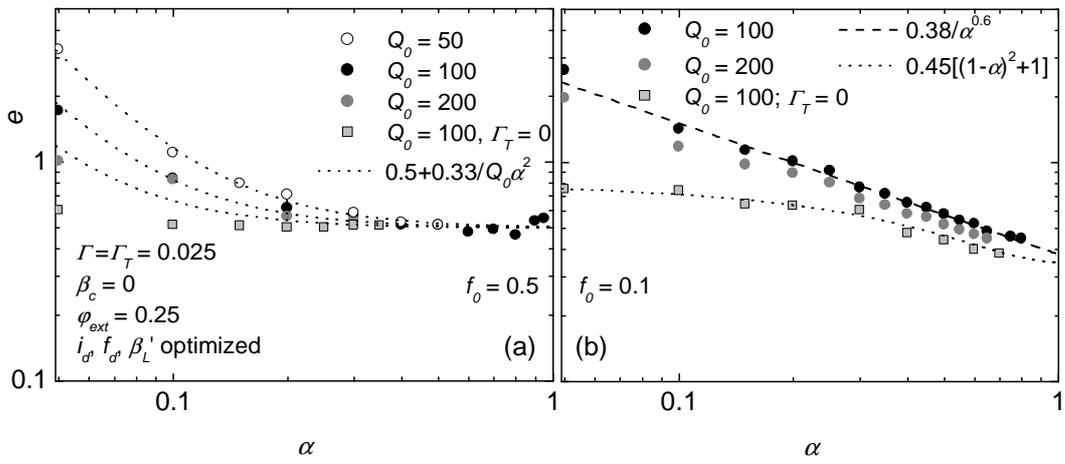

**Figure 11**

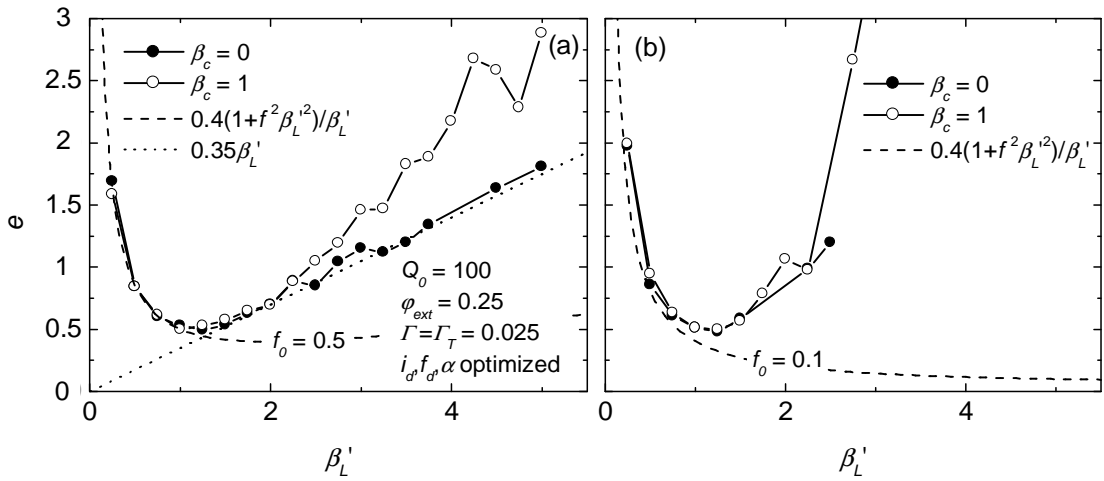

**Figure 12**

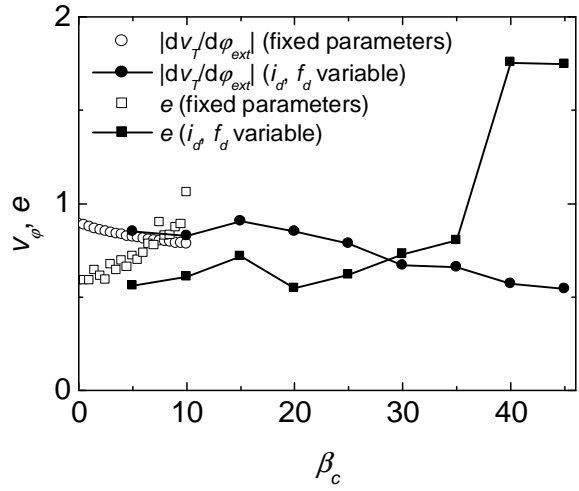

**Figure 13**

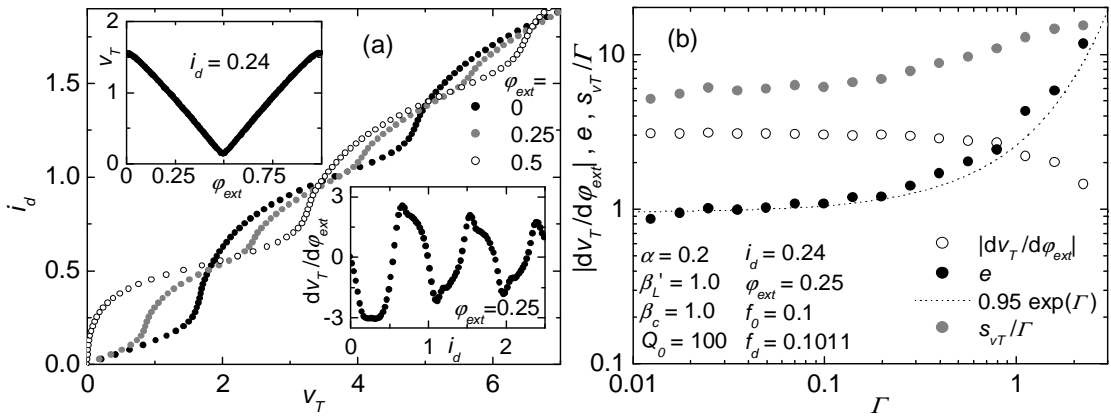

**Figure 14**

# Tables

| $f_0$ | $i_d$ | $\xi$ | $\alpha$ | $\beta'_L$ | $dv_T/d\varphi_{ext}$ | $s_{vT}$ | $e$ |
|---|---|---|---|---|---|---|---|
| 1 | 0.444 | 1.25 | 0.194 | 1.103 | -23.7 | 5.8 | 0.58 |
| 0.5 | 0.338 | 1 | 0.231 | 1.247 | -12.4 | 1.84 | 0.6 |
| 0.333 | 0.331 | 4.5 | 0.413 | 1.25 | -5.14 | 0.3 | 0.58 |
| 0.25 | 0.263 | 3.5 | 0.313 | 1.075 | -5.45 | 0.26 | 0.52 |
| 0.2 | 0.219 | 2.5 | 0.250 | 1.031 | -5.26 | 0.26 | 0.56 |
| 0.167 | 0.244 | 3.75 | 0.334 | 0.975 | -3.49 | 0.104 | 0.56 |
| 0.143 | 0.388 | 12.4 | 0.625 | 1.116 | -1.69 | 0.024 | 0.48 |
| 0.1 | 0.369 | 13.2 | 0.725 | 1.213 | -1.08 | 0.012 | 0.48 |
| 0.0333 | 0.281 | 9.2 | 0.706 | 1.05 | -0.43 | 0.00185 | 0.6 |
| 0.01 | 0.3 | 9.8 | 0.772 | 1 | -0.127 | 0.00016 | 0.63 |

**Table 1**

| $\alpha$ | $f_0$ | $Q_0$ | $\Gamma_T$ | $i_d$ | $\xi$ | $\beta'_L$ | $dv_T/d\varphi_{ext}$ | $e$ |
|---|---|---|---|---|---|---|---|---|
| 0.05 | 0.5 | 100 | 0.025 | 0.44 | -0.11 | 1.56 | -18.2 | 1.7 |
| 0.2 | 0.5 | 100 | 0.025 | 0.29 | 1.25 | 1.1 | -14.5 | 0.6 |
| 0.5 | 0.5 | 100 | 0.025 | 0.58 | 10 | 1.2 | -6.5 | 0.5 |
| 0.7 | 0.5 | 100 | 0.025 | 0.98 | 29.5 | 1.36 | -4.4 | 0.5 |
| 0.9 | 0.5 | 100 | 0.025 | 1.5 | 45.5 | 1.24 | -2.9 | 0.55 |
| 0.05 | 0.5 | 200 | 0.025 | 0.24 | 0.14 | 1.39 | -29.9 | 1.0 |
| 0.2 | 0.5 | 200 | 0.025 | 0.24 | 2.3 | 1.18 | -15.8 | 0.55 |
| 0.05 | 0.5 | 100 | 0 | 0.5 | 0.13 | 1.05 | -15.4 | 0.6 |
| 0.2 | 0.5 | 100 | 0 | 0.3 | 0.94 | 1.05 | -14.0 | 0.5 |
| 0.3 | 0.5 | 100 | 0 | 0.38 | 1.7 | 1.13 | -9.7 | 0.5 |
| 0.05 | 0.1 | 100 | 0.025 | 0.34 | 0.7 | 1.525 | -6.0 | 2.6 |
| 0.2 | 0.1 | 100 | 0.025 | 0.19 | 2.1 | 1.1 | -3.15 | 1.1 |
| 0.5 | 0.1 | 100 | 0.025 | 0.29 | 8.8 | 1.11 | -1.45 | 0.6 |
| 0.75 | 0.1 | 100 | 0.025 | 0.44 | 18.4 | 1.27 | -1.0 | 0.45 |
| 0.05 | 0.1 | 200 | 0.025 | 0.32 | 1.9 | 1.31 | -6.1 | 2.0 |
| 0.2 | 0.1 | 200 | 0.025 | 0.15 | 3.7 | 1.08 | -3.4 | 0.9 |
| 0.5 | 0.1 | 200 | 0.025 | 0.34 | 18.4 | 1.31 | -1.5 | 0.5 |
| 0.05 | 0.1 | 100 | 0 | 0.41 | 0.5 | 1.08 | -3.4 | 0.75 |
| 0.2 | 0.1 | 100 | 0 | 0.24 | 2.1 | 0.85 | -3.0 | 0.65 |
| 0.5 | 0.1 | 100 | 0 | 0.35 | 8.2 | 0.88 | -1.55 | 0.45 |
| 0.7 | 0.1 | 100 | 0 | 0.41 | 12.7 | 1.01 | -1.15 | 0.4 |

**Table 2**

| $\beta'_L$ | $\beta_c$ | $f_0$ | $i_d$ | $\xi$ | $\alpha$ | $dv_T/d\varphi_{ext}$ | $e$ |
|---|---|---|---|---|---|---|---|
| 0.25 | 0 | 0.5 | 1.22 | 7.77 | 0.51 | -4.6 | 1.7 |
| 0.5 | 0 | 0.5 | 0.77 | 8.24 | 0.46 | -6.3 | 0.85 |
| 0.75 | 0 | 0.5 | 0.7 | 10.7 | 0.45 | -6.8 | 0.6 |
| 1 | 0 | 0.5 | 0.515 | 8.0 | 0.42 | -8.0 | 0.55 |
| 1.25 | 0 | 0.5 | 0.58 | 10.5 | 0.52 | -6.1 | 0.5 |
| 1.5 | 0 | 0.5 | 0.53 | 6.5 | 0.47 | -6.7 | 0.55 |
| 2 | 0 | 0.5 | 0.45 | 3.7 | 0.38 | -7.8 | 0.7 |
| 3 | 0 | 0.5 | 0.28 | 2.7 | 0.24 | -13.4 | 1.15 |
| 0.25 | 1 | 0.5 | 1.5 | 10.4 | 0.62 | -3.6 | 1.6 |
| 0.5 | 1 | 0.5 | 0.73 | 5.37 | 0.41 | -6.4 | 0.85 |
| 0.75 | 1 | 0.5 | 0.53 | 3.3 | 0.36 | -7.9 | 0.6 |
| 1 | 1 | 0.5 | 0.75 | 6.8 | 0.53 | -5.5 | 0.5 |
| 1.25 | 1 | 0.5 | 0.58 | -0.82 | 0.33 | -8.1 | 0.55 |
| 1.5 | 1 | 0.5 | 0.46 | -1.26 | 0.29 | -11.0 | 0.6 |
| 2 | 1 | 0.5 | 0.67 | -2.83 | 0.33 | -8.0 | 0.7 |
| 3 | 1 | 0.5 | 0.28 | 3.79 | 0.30 | -9.0 | 1.45 |
| 0.25 | 0 | 0.1 | 1.05 | 12.5 | 0.94 | -1.0 | 2.0 |
| 0.5 | 0 | 0.1 | 1.05 | 26.5 | 0.95 | -0.85 | 0.85 |
| 0.75 | 0 | 0.1 | 0.8 | 23.5 | 0.82 | -0.97 | 0.6 |
| 1 | 0 | 0.1 | 0.63 | 24.25 | 0.83 | -0.94 | 0.5 |
| 1.25 | 0 | 0.1 | 0.61 | 25.75 | 0.83 | -0.92 | 0.5 |
| 1.5 | 0 | 0.1 | 0.49 | 22.75 | 0.83 | -0.89 | 0.6 |
| 2.5 | 0 | 0.1 | 0.74 | 18.25 | 0.84 | 0.81 | 1.2 |
| 0.25 | 1 | 0.1 | 1.05 | 12.5 | 0.93 | -1.0 | 2.0 |
| 0.5 | 1 | 0.1 | 0.73 | 13 | 0.78 | -1.2 | 0.95 |
| 0.75 | 1 | 0.1 | 0.73 | 19.5 | 0.78 | -1.05 | 0.65 |
| 1 | 1 | 0.1 | 0.7 | 26.5 | 0.81 | -0.92 | 0.5 |
| 1.25 | 1 | 0.1 | 0.65 | 27.5 | 0.8 | -0.91 | 0.5 |
| 1.5 | 1 | 0.1 | 0.53 | 22.5 | 0.79 | -0.92 | 0.55 |
| 2.25 | 1 | 0.1 | 0.93 | 23.6 | 0.9 | 0.83 | 1.0 |
| 3 | 1 | 0.1 | 0.31 | 26.1 | 0.93 | -1.2 | 3.5 |

**Table 3**